\documentclass[sigconf]{acmart}

\AtBeginDocument{%
  
}

\usepackage[utf8]{inputenc}
\usepackage{microtype}
\usepackage{multirow}
\usepackage{subfigure}
\usepackage[utf8]{inputenc}
\usepackage{graphicx}
\usepackage{amsmath}
\usepackage{algorithm}
\usepackage{algorithmic}
\usepackage{CJKutf8}
\usepackage{balance}
\usepackage[skip=2.5pt]{caption}
\usepackage{array}
\usepackage{booktabs}
\usepackage{threeparttable}
\usepackage{soul}
\usepackage{setspace}

\setcopyright{acmlicensed}
\copyrightyear{2018}
\acmYear{2018}
\acmDOI{XXXXXXX.XXXXXXX}

\acmConference[Conference acronym 'XX]{Make sure to enter the correct
  conference title from your rights confirmation emai}{June 03--05,
  2018}{Woodstock, NY}
\acmISBN{978-1-4503-XXXX-X/18/06}

\copyrightyear{2025}
\acmYear{2025}
\setcopyright{acmlicensed}\acmConference[WWW Companion '25]{Companion Proceedings of the ACM Web Conference 2025}{April 28-May 2, 2025}{Sydney, NSW, Australia}
\acmBooktitle{Companion Proceedings of the ACM Web Conference 2025 (WWW Companion '25), April 28-May 2, 2025, Sydney, NSW, Australia}
\acmDOI{10.1145/3701716.3715228}
\acmISBN{979-8-4007-1331-6/25/04}

\begin{document}

%%
%% The "title" command has an optional parameter,
%% allowing the author to define a "short title" to be used in page headers.
\title{
Generative Retrieval and Alignment Model: A New Paradigm for E-commerce Retrieval
}

%%
%% The "author" command and its associated commands are used to define
%% the authors and their affiliations.
%% Of note is the shared affiliation of the first two authors, and the
%% "authornote" and "authornotemark" commands
%% used to denote shared contribution to the research.

\author{Ming Pang}
\authornote{Both authors contributed equally to this research.}
\authornote{is corresponding author.}
\email{pangming8@jd.com}
\author{Chunyuan Yuan}
\email{chunyuany93@outlook.com}
\authornotemark[1]
\affiliation{
  \institution{JD.COM}
  \city{Beijing}
  \country{China}
}

\author{Xiaoyu He}
\email{hexiaoyu5@jd.com}
% \authornote{Both authors are corresponding authors.}
\author{Zheng Fang}
\email{fangzheng21@jd.com}
% \authornotemark[2]
\affiliation{
  \institution{JD.COM}
  \city{Beijing}
  \country{China}
}

\author{Donghao Xie}
\email{xiedonghao1@jd.com}
\author{Fanyi Qu}
\email{qufanyi1@jd.com}
\affiliation{
  \institution{JD.COM}
  \city{Beijing}
  \country{China}
}

% \author{Hao Jiang}
% \email{jianghao163@jd.com}
% \author{Yujing Lin}
% \email{linyujing5@jd.com}
% \affiliation{
%   \institution{JD.COM}
%   \city{Beijing}
%   \country{China}
% }

\author{Xue Jiang}
\email{jiangxue@jd.com}
\author{Changping Peng}
\email{pengchangping@jd.com}
\affiliation{
  \institution{JD.COM}
  \city{Beijing}
  \country{China}
}

\author{Zhangang Lin}
\email{linzhangang@jd.com}
\author{Ching Law}
\email{lawching@jd.com}
\affiliation{
  \institution{JD.COM}
  \city{Beijing}
  \country{China}
}

\author{Jingping Shao}
\email{shaojingping@jd.com}
\affiliation{
  \institution{JD.COM}
  \city{Beijing}
  \country{China}
}

%%
%% By default, the full list of authors will be used in the page
%% headers. Often, this list is too long, and will overlap
%% other information printed in the page headers. This command allows
%% the author to define a more concise list
%% of authors' names for this purpose.
\renewcommand{\shortauthors}{Ming Pang et al.}

%%
%% The abstract is a short summary of the work to be presented in the
%% article.
\begin{abstract}
The retrieval module is a crucial component of search systems. Traditional sparse and dense retrieval methods struggle to leverage general world knowledge and often fail to capture the nuanced features of queries and products. With the advent of large language models (LLMs), industrial search systems have started to employ LLMs to generate identifiers for product retrieval. Commonly used identifiers include (1) static/semantic IDs and (2) product term sets. The first approach requires creating a product ID system from scratch, missing out on the world knowledge embedded within LLMs. While the second approach leverages this general knowledge, the significant difference in word distribution between queries and products means that product-based identifiers often do not align well with user search queries, leading to missed product recalls. Furthermore, when queries contain numerous attributes, these algorithms generate a large number of identifiers, making it difficult to assess their quality, which results in low overall recall efficiency.

To address these challenges, this paper introduces a novel e-commerce retrieval paradigm: the Generative Retrieval and Alignment Model (GRAM). GRAM employs joint training on text information from both queries and products to generate shared text identifier codes, effectively bridging the gap between queries and products. This approach not only enhances the connection between queries and products but also improves inference efficiency. The model uses a co-alignment strategy to generate codes optimized for maximizing retrieval efficiency. Additionally, it introduces a query-product scoring mechanism to compare product values across different codes, further boosting retrieval efficiency. By integrating these scores, GRAM can replace the traditional recall and initial ranking stages, achieving an integrated retrieval and pre-ranking process. Extensive offline and online A/B testing was conducted. The results demonstrate that GRAM significantly outperforms traditional sparse and dense retrieval algorithms and the latest generative retrieval models, confirming its effectiveness and practicality.
\end{abstract}

%%
%% The code below is generated by the tool at http://dl.acm.org/ccs.cfm.
%% Please copy and paste the code instead of the example below.
%%

%%
%% Keywords. The author(s) should pick words that accurately describe
%% the work being presented. Separate the keywords with commas.

%% A "teaser" image appears between the author and affiliation
%% information and the body of the document, and typically spans the
%% page.
% \begin{teaserfigure}
%   \includegraphics[width=\textwidth]{sampleteaser}
%   \caption{Seattle Mariners at Spring Training, 2010.}
%   \Description{Enjoying the baseball game from the third-base
%   seats. Ichiro Suzuki preparing to bat.}
%   \label{fig:teaser}
% \end{teaserfigure}

% \received{20 February 2007}
% \received[revised]{12 March 2009}
% \received[accepted]{5 June 2009}

%%
%% This command processes the author and affiliation and title
%% information and builds the first part of the formatted document.
\maketitle

\section{Introduction}
Online shopping has greatly changed our lives and has become an indispensable part of daily life. In the past few years, more and more e-commerce platforms such as Amazon, JD.com, and Taobao have provided consumers with hundreds of millions of colorful products. How to facilitate consumers to quickly and accurately retrieve the products they need from these massive products has become an extremely important research topic.

Traditional e-commerce retrieval models~\cite{robertson1994some,dean2009challenges} use sparse retrieval methods such as inverted indexes~\cite{dean2009challenges} to retrieve products. With the development of representation learning~\cite{bengio2013representation} and deep pre-trained language models~\cite{schmidhuber2015deep,kenton2019bert,liu2019roberta}, dense retrieval has been widely used in online e-commerce retrieval systems. Dense retrieval~\cite{karpukhin2020dense,yuan2023multi,yuan2024semi} maps query and product texts to the same vector space for matching. 
Compared to sparse retrieval, dense retrieval offers improved semantic matching and significantly increases efficiency. However, it has several limitations. First, there is a lack of deep interaction between queries and products, making it difficult to represent fine-grained information, which limits retrieval effectiveness. Second, in industrial scenarios with large product scales, storing representations demands substantial memory. Lastly, model training is constrained by the construction of positive and negative samples, preventing the use of general knowledge of the world.

With the advancement of large-language models, there is growing interest in industrial search systems to explore these models for enhancing the efficiency of their retrieval modules. However, the industry has not yet widely adopted LLM-based generative retrieval. A central challenge is the selection of identifier codes which can be categorized into two main types:
\begin{itemize}
\item ID-Based document codes~\cite{rajput2023recommender,yang2023auto,zheng2024adapting,zeng2024scalable}: This category includes both static and semantic IDs. Static IDs~\cite{zeng2024scalable} require the model to directly generate document or product IDs, which are then used to recall the corresponding items. Semantic IDs~\cite{rajput2023recommender,yang2023auto,zheng2024adapting} involve semantically representing the text or product to be recalled, assigning a unique ID to each recalled item through multi-level semantic clustering. While this approach effectively maps a product or document to a unique identifier, it poses significant challenges in terms of generation accuracy and recall efficiency. For instance, recalling $k$ items necessitates the use of beam search to return $k$ results, resulting in a time complexity that can be approximated as $O(k^2)$. Furthermore, this approach necessitates learning a new ID system from scratch, which limits the use of the extensive general knowledge embedded in LLMs.

\item String-based document codes~\cite{bevilacqua2022autoregressive,lee2023glen,li2024distillation}: This approach includes sub-strings, which extract n-grams from text as representations, and term sets, which utilize keyword collections for representation. Although these algorithms capitalize on the general knowledge of LLMs, they construct query and product identifiers independently. The significant differences in word distribution between the two text types can severely impact overall recall efficiency. Moreover, when the number of attributes in the query is large, these algorithms can generate excessive identifiers. The quality of these identifiers is often difficult to assess, further diminishing the recall efficiency of the online system.
\end{itemize}

To address the above challenges, this paper proposes a new e-commerce retrieval paradigm: the Generative Retrieval and Alignment Model (GRAM). We construct product codes based on NER attributes and use a large model to generate query and product identifier codes. These attributes, organized in natural language, fully leverage the general world knowledge embedded in LLM pre-training. By employing joint training on text information from both queries and products, we generate shared text codes, thereby resolving issues of inconsistent codes caused by differences in word distribution between query and product texts, and enhancing the efficiency of code generation and retrieval. We further assess code quality by examining the variation in recall efficiency of products retrieved via these codes, using a preference alignment algorithm to increase the likelihood of generating high-quality codes, thus boosting retrieval efficiency.

The contributions of this paper are as follows:
\begin{itemize}
\item We introduce an innovative and practical approach that utilizes the inherent world knowledge of LLMs to generate shared codes for queries and products, effectively overcoming the inconsistency issues due to differing word distributions in the two domains.
\item We develop the GRAM model, which effectively differentiates the quality of codes based on the variation in product recall efficiency during code retrieval, thereby enhancing the model’s ability to generate high-quality codes and improve retrieval efficiency. Additionally, by introducing a query-product scoring mechanism, GRAM can simultaneously replace traditional recall and initial ranking stages, achieving an integrated retrieval and pre-ranking process and further boosting retrieval efficiency.
\item The effectiveness of GRAM has been validated through extensive offline experiments on a large-scale real-world dataset and online A/B testing. GRAM significantly outperforms traditional sparse and dense retrieval algorithms, as well as the latest generative retrieval models in the industry, confirming its effectiveness and practicality. It has been deployed on the JD e-commerce platform, providing hundreds of millions of product retrieval services every day, showcasing its high commercial value and serving as a practical and robust large-scale product retrieval solution.
\end{itemize}

\section{Related Work}
Research on document retrieval can be broadly categorized into three distinct types: sparse retrieval, dense retrieval, and generative retrieval. A brief overview of each category is provided below.

\subsection{Sparse Retrieval}
Sparse retrieval techniques are fundamental to traditional information retrieval, utilizing inverted indexing to map unique terms to documents efficiently. This method allows for rapid access to relevant information in large collections. A key metric in this domain is the Term Frequency-Inverse Document Frequency (TF-IDF)~\cite{ramos2003using}, which assesses the significance of terms across documents and is widely used in retrieval systems. Early research primarily focused on inverted indexing with term-matching metrics such as TF-IDF. BM25~\cite{robertson2009probabilistic} enhanced relevance scoring by refining term weights based on the TF-IDF feature. Recent studies~\cite{zheng2015learning,dai2020context} have integrated word embeddings into inverted indexing to address the problem of term mismatch. 

Despite the effectiveness of sparse retrieval methods in delivering fast results, they still struggle with intricate queries involving synonyms, specialized terms, or contextual nuances, underscoring the need for ongoing advancements to meet users' diverse information needs better.

\subsection{Dense Retrieval}
The fundamental principle of dense retrieval is to transform documents and queries into vector representations. The introduction of pre-trained language models, particularly BERT~\cite{kenton2019bert}, has revolutionized information retrieval, paving the way for dense retrieval methods such as Dense Passage Retrieval (DPR)~\cite{karpukhin2020dense}, ColBERT~\cite{khattab2020colbert}, and GTR~\cite{ni2022large}. Techniques like SimCSE~\cite{gao2021simcse} leverage contrastive learning with models such as BERT and Roberta to optimize embeddings. Additionally, dense retrieval methods often employ Approximate Nearest Neighbor (ANN) search~\cite{jayaram2019diskann,xiongapproximate}, Maximum Inner Product Search (MIPS) algorithms~\cite{shrivastava2014asymmetric}, and SimLM~\cite{wang2023simlm} to ensure efficient retrieval in sub-linear time. 

Unlike traditional sparse retrieval, these methods utilize transformer encoders to create dense vector representations for queries and documents, enhancing semantic understanding and retrieval accuracy. This combination of semantic depth and computational efficiency positions dense retrieval as a leading approach in modern information retrieval.

\subsection{Generative Retrieval}
Generative retrieval has emerged as a promising paradigm in information retrieval, leveraging the advancements in pre-trained models to generate document identifiers (DocIDs) from user queries directly, thus eliminating the need for traditional index-based methods. Early works~\cite{tay2022transformer} in this area introduced transformer auto-regressive models that preprocess documents into atomic or hierarchical identifiers using hierarchical k-means clustering. Other approaches, like SEAL~\cite{bevilacqua2022autoregressive}, utilized n-grams as identifiers. Ultron~\cite{zhou2022ultron} combines keyword-based and semantic-based identifiers within a three-stage training process. TIGER~\cite{rajput2023recommender} utilizes semantic IDs for product indexing and generates the semantic ID of the next item. Additionally, SE-DSI~\cite{tang2023semantic} proposed using summarization texts as document identifiers, and RIPOR~\cite{zeng2024scalable} uses an encoder-decoder model as the backbone, where a dense encoder encodes document content and a decoder uses a start token for decoding. GLEN~\cite{lee2023glen} designs dynamic lexical DocIDs and is trained through a two-phase index learning strategy.

These generative retrieval models offer greater flexibility and semantic understanding, enabling end-to-end optimization and reducing dependence on external indexing. However, they either require building product ID systems from scratch, failing to fully utilize the world knowledge embedded in LLMs, or face difficulties in aligning product-based identifiers with user search queries due to significant differences in word distribution. Additionally, when queries have many attributes, these algorithms produce a large number of identifiers, making it challenging to assess their quality, which leads to low overall recall efficiency. Moreover, the generation complexity of the ID-based methods is linear to the beam size, which makes them impractical for industrial applications.

\section{Model}
In this section, we begin by formally defining the generative retrieval task. Following that, we provide a detailed description of the various modules within GRAM and examine the model’s impact during both the training and inference phases.

\begin{figure*}[!htbp]
    \centering
    \includegraphics[scale=1.0]{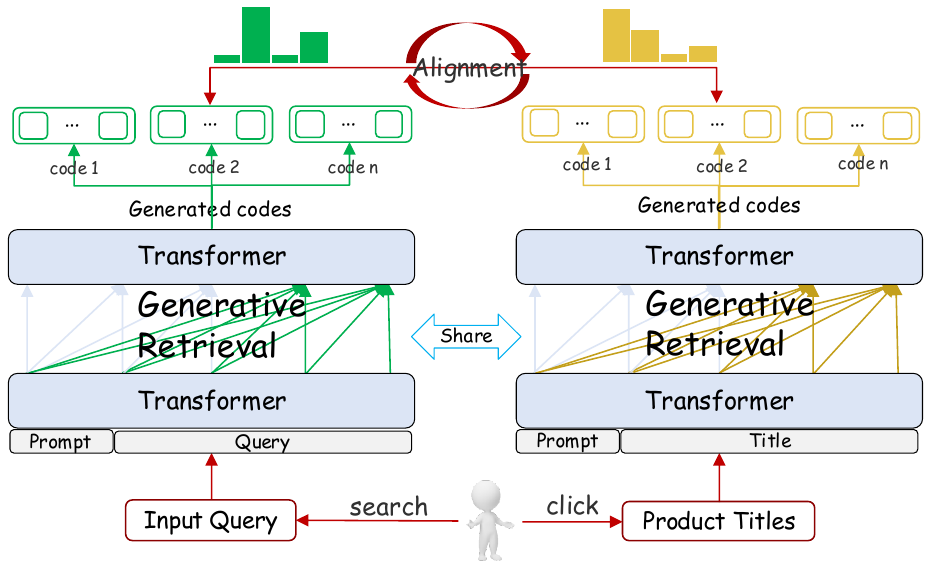}
    \caption{The architecture of the generative retrieval and alignment model.}
    \label{model_structure}
\end{figure*}

\subsection{Problem Statement}
Suppose the query inputted by users on the E-commerce applications, has $q = [q_1, q_2, \ldots, q_{L_q}]$ characters. After browsing the search result list, the user clicks product with title $t = [t_1, t_2, \ldots, t_{L_t}]$. 

Our target is to train a generative retrieval model $f(c | x, \pi_\theta), x \in \{q_i, t_j\}$ to generate product code for any query $q_i$ or product title $t_j$ and ensure that the input query $q$ can retrieve the relevant product $t$ by the code $c$. Furthermore, considering the retrieval efficiency, we should optimize the generation process to ensure the code $c$ can retrieve as many relevant products for query $q$ as possible. For a clear definition, bold lowercase letters represent vectors throughout the rest of this paper.

\subsection{Overview}
Figure~\ref{model_structure} illustrates the components of the GRAM, which comprises three primary modules: (1) the query-code generator, (2) the product-code generator, and (3) the code co-alignment module. Specifically, the query-code generator takes a query as input and generates a series of predefined text identifiers, which serve as codes that bridge the connection between the query and products. Similarly, the product-code generator uses the product title as input to generate codes that denote the product itself. The code co-alignment module is designed to ensure that the codes generated by both the query and its associated products align effectively, thereby maximizing the overall retrieval efficiency of the search system.

\subsection{Code Definition and Construction}
In this subsection, we will formally define the query and product code composition and how to construct the initial codes to drive the GRAM training process. 

\subsubsection{Code Definition}
Previous research has examined various coding approaches based on identifiers (IDs)~\cite{rajput2023recommender,yang2023auto,zheng2024adapting,zeng2024scalable} and strings~\cite{bevilacqua2022autoregressive,lee2023glen,li2024distillation}. ID-based methods require training product ID systems from scratch, which limits their ability to take advantage of the extensive world knowledge embedded in large language models (LLMs). This limitation significantly undermines the potential effectiveness of LLMs. Therefore, we propose using strings as a coding mechanism to represent products and articulate the intent behind user queries.

In contrast to previous studies~\cite{bevilacqua2022autoregressive,lee2023glen,li2024distillation}, we do not employ substrings or n-grams derived from queries or product titles as codes. Instead, we focus on the structural attributes of both queries and products to accurately convey user intent and product characteristics, as both are fundamentally expressed through these structural attributes. Specifically, we have identified 16 commonly used structured attributes as the foundational elements of our coding system: brand names, product categories, series, models, functional attributes, material attributes, style attributes, color attributes, sales specifications, technical specifications, applicable time, product audiences, applicable scenarios, additional modifiers, and marketing terminology. Using these structural attributes enables us to effectively describe nearly any product within the e-commerce system.

The coding system is constructed from one or more structured attributes, which are delineated by separators. We define the granularity of the code as follows:
\begin{itemize}
\item Coarse-grained code: consists of 1-2 structured attributes connected by delimiters;
\item Medium-grained code: consists of 3 structured attributes connected by delimiters;
\item Fine-grained code: consists of more than 3 structured attributes connected by delimiters.
\end{itemize}

\subsubsection{Initial Code Construction}
First, we use experts to annotate the query with the predefined key attributes above. The results of the two-person labeling are used as reliable labeling samples. The scale of the labeled data is about 3 million. Secondly, we used these labeled data to train a BERT-based NER model. Subsequently, we use the NER discriminant model to extract the key attributes defined above from the query to form a set of key attributes $Q$. Then, we mine the high-frequency query-product pair click data based on the query click log. At the same time, we use the NER discriminant model to extract the corresponding attributes from the titles of these products. It is necessary to ensure that the extracted attributes are in the set $Q$. 

The initial code of the query consists of two parts: (1) the key attribute combination extracted by the query through the NER model; and (2) the key attribute combination extracted by the product associated with the query through the click relationship. The initial code of the product consists of two parts: (1) the key attribute combination extracted by the product through NER; and (2) the key attribute combination extracted by the query that is reversely associated with the product through the click relationship. The initial query-code pair and product-code pair data constructed in this way serve as the GRAM's initial training data, driving the entire model's interactive training. The initial training dataset contains about 6 million unique queries and 8 million unique products extracted from the search log of e-commerce applications.

\subsection{Supervised Fine-Tuning}
In this section, we will outline the process of training the query-code generator and the product-code generator separately, as well as how to utilize these generators for product retrieval. In addition, we will discuss the iterative process of integrating the two generators to enhance the generation of code for both components.

The primary goal of supervised fine-tuning (SFT) training at this stage is to improve the retrieval rate of products retrieved through query-based code generation.

\subsubsection{Query-code generator SFT}
The query-code generator is trained by the query-code pair data constructed above, with the query text as input and the code text as the target. The open-source LLM is used as the base model for further training. To further improve the generation performance of the model for e-commerce text, we introduce task-specific supervised fine-tuning (SFT). The process of generating text with a conditional language model can be conceptualized as a constrained auto-regressive sampling strategy. Given a query $q$ and its corresponding gold standard code $c$, we design a special prompt text $Pmpt_q$ to instruct training, the training objective is to maximize the conditional probability $Pr(c|Pmpt_q, q)$. Specifically, the training objective for the query-code generation model involves minimizing the negative log-likelihood, which can be formulated as follows: 
\begin{equation}
\begin{split}
& \mathcal{L}^q_\mathrm{SFT} = - \mathbb{E}_{(q, c) \sim \mathcal{D}^q_\mathrm{SFT}} \sum_{i=1}^{L_c} \log Pr(c_i | Pmpt_q, q, c_{<i})  \,, \\ 
\end{split}
\end{equation}
where $c_i$ is the $i$-th token of the code $c$, $D^q_\mathrm{SFT}$ is the collected query-code training data, $L_c$ is the length of the query code.

\subsubsection{Product-code generator SFT}
The product-code generator is trained using the product-code pair data constructed above, with the product title text as input and the code text as the target. The process of generating text with a conditional language model can be conceptualized as a constrained auto-regressive sampling strategy. Given a product $t$ and its corresponding gold standard code $c$, we design a special product prompt $Pmpt_t$ to instruct GRAM training, the objective is to maximize the conditional probability $Pr(c|Pmpt_t, t)$. Specifically, the training objective for the product-code generator model involves minimizing the negative log-likelihood, which can be formulated as follows: 
\begin{equation}
\begin{split}
& \mathcal{L}^t_\mathrm{SFT} = - \mathbb{E}_{(t, c) \sim \mathcal{D}^t_\mathrm{SFT}} \sum_{i=1}^{L_c} \log Pr(c_i | Pmpt_t, t, c_{<i})  \,, \\ 
\end{split}
\end{equation}
where $c_i$ is the $i$-th token of the code $c$, $D^t_\mathrm{SFT}$ is the collected product-code training data, $L_c$ is the length of the product code.

\subsubsection{Generator Co-training}
Through the training process outlined above, the GRAM acquires fundamental capabilities for generating query-to-code and product-to-code mappings. We randomly selected a batch of queries and products to generate their corresponding codes. Due to the limited diversity of queries and the small number of products represented in the training data, the model predominantly generated fine-grained codes for both queries and products. Although these fine-grained codes improve the relevance of retrieved products, they can impede retrieval efficiency, as many mid- and long-tail queries and products may not be effectively captured by the generated codes.

To address these issues, we utilized the trained product-code generator to create code data for the active products within the e-commerce system. Simultaneously, leveraging the co-click relationship, we fed the queries associated with these products into the query-code generator, generating 10 new codes for each query. After deduplicating the newly generated codes, they were evaluated using a trained query-code and product-code relevance model. The codes that passed relevance filtering were incorporated into the existing set of codes as augmented training data.

Unlike the separate training conducted in the previous stages, this phase needs to strengthen the connection between the two models, thus training the two generators simultaneously. The objective function for this training is formulated as follows:
\begin{equation}
\begin{split}
& \mathcal{L}_\mathrm{SFT} =  \mathcal{L}^q_\mathrm{SFT} + \lambda * \mathcal{L}^t_\mathrm{SFT} \,, \\ 
\end{split}
\end{equation}
where $\lambda$ is a hyper-parameter to adjust the weight of the loss function for the title-code generator.

\subsection{Query-Code-Product Co-Alignment}
In this section, we will describe the process of aligning the codes generated by the query code generator with those produced by the product code generator. This alignment is intended to enhance the retrieval rate of products through query-based code generation. Additionally, we will discuss how to achieve preference alignment between queries and products within the same code framework, which will contribute to an increased proportion of recalled products that successfully pass through the query-product relevance filtering module.

The goal of co-alignment is to maximize the proportion of retrieved products that pass the query-product relevance filtering module while minimizing any potential decrease in the recall rate of products retrieved through query-based code generation.

\subsubsection{Codes Co-Alignment}
To achieve the overall goal of this phase, the first step is to align the codes generated by the query-code generator with those produced by the product-code generator. Firstly, we take users' click data from the search log to build a query-product pair dataset. Subsequently, we take the query-code and product-code corresponding to each query-product pair from the initial query-code and product-code training dataset. Finally, we take the intersection of the two datasets as the positive example set, and the difference between the two as the negative example set to build a partial-order dataset $\mathcal{D}_{CA}$.

There are two main considerations why we directly use the query-code pair and product-code pair from the initial training data. The first reason is to align with the initial version of the model and directly correct the results of the model SFT based on the original data; the other is that there are more low-quality codes (especially on the query side) in the initial data, and the probability of selecting pseudo-negative examples during random selection is lower overall.

A single training sample consists of a query $q$, product $t$, positive code $c_w$, and negative code $c_l$. Direct preference optimization algorithm~\cite{rafailov2024direct} is used for code alignment. 
The GRAM, which is trained in the supervised fine-tuning process, is used as the reference model $\pi_\mathrm{SFT}(\cdot)$ to calculate the probability values. 
\begin{equation}
\begin{split}
& \pi_{SFT}(c_w|q,t) = \left(\pi_{SFT}(c_w| q) + \pi_{SFT}(c_w|t) \right) / 2   \,,  \\
& \pi_{SFT}(c_l|q,t) = \left(\pi_{SFT}(c_l| q) + \pi_{SFT}(c_l|t) \right) / 2   \,.  \\
\end{split}
\end{equation}

Similarly,  $\pi_\mathrm{\theta}(\cdot)$ is applied to calculate the generation probability of positive code and negative code under query, and the generation probability of positive code and negative code under products. Specifically, the process can be formulated as follows: 
\begin{equation}
\begin{split}
& \pi_{\theta}(c_w|q,t) = \left(\pi_{\theta}(c_w| q) + \pi_{\theta}(c_w|t) \right) / 2   \,,  \\
& \pi_{\theta}(c_l|q,t) = \left(\pi_{\theta}(c_l| q) + \pi_{\theta}(c_l|t) \right) / 2   \,.  \\
\end{split}
\end{equation}

After obtaining the probabilities, the overall training objective can be formulated as follows:
\begin{align}
\mathcal{L}_{CA} (\pi_{\theta}) = & -\mathbb{E}_{(q, t, c_w, c_l) \sim \mathcal{D}_{CA}} \left[ \log\sigma\left( \beta_{w} \log\frac{\pi_{\theta}(c_w | q, t)}{\pi_\mathrm{SFT}(c_w | q, t)} \right. \right. \nonumber \\  & \left. \left.  - \beta_l \log\frac{\pi_{\theta}(c_l | q, t)}{\pi_\mathrm{SFT}(c_l | q, t)} \right) \right]   \,.  \\ \nonumber 
\end{align}

In the experiment, we found that the probability of fine-grained codes being selected as negative examples is higher than that of coarse-grained codes, and the probability of coarse-grained codes being selected as positive examples is also higher than that of fine-grained codes. This bias may be reasonable for a single piece of data, but it will cause the generated results to be more biased towards coarse-grained codes overall. Therefore, we perform a second data downsampling. For all positive codes matched under the query-product pair, we add a length penalty factor $\alpha = \frac{code_l}{max_{code_l}}$. Finally, the number of repetitions of a positive code in the training data is $n=sqrt(k) * \alpha$.

\subsubsection{Query-product Alignment through Code}
The above data construction method and model training method use query-product as anchor points and select positive and negative example codes for preference alignment training. In this way, the model can generate a higher probability for codes with display/click behaviors than the codes without display/click behaviors, which is more about learning the partial order relations of code. However, the partial orders between the query and product under the code are still unclear, so further model alignment is needed.

Based on the model trained in the previous step, recall the product set for any query-code pair. Use the online correlation results to filter out positive and negative products from the recalled product set. Based on the query-code results and product-code results, complete the query-product recall through code matching. For such a set of (query, code, product) triples, calculate the query-product score by token based on the Jensen-Shannon divergence. 

For the same token of code, we use the distance between the probability of the query and the product generating this token as the basis for scoring. The process can be formulated as follows:
\begin{equation}
S_{rele}(q, t) = \sum_{i=1}^{n} w_i * \left(P_i^q * ln\frac{2P_i^q}{P_i^q + P_i^t} + P_i^t *  ln\frac{2P_i^t}{P_i^q + P_i^t} \right)  \,, \\
\end{equation}
where $P_i^q$ is the probability of generating the $i$-th token in the query generated code is a scalar between 0 and 1. $P_i^t$ is the probability of generating the $i$-th token in the product generation code. $w_i$ is code-granular, and each code has a weight. The model parameters are fixed and only the weights are trained.

A product may be retrieved by multiple codes of the same query. Each code will have a relevance score $S_{rele}$, and the scores of all codes need to be summed up as the relevance of the product.

\subsubsection{Overall Alignment Objective} 
After obtaining the query-product scores through the above process, we construct a partial order relationship of the products retrieved by the query through the code. At the same time, each query uses the top k products with the highest scores as positive samples $t_{pos}$ and the other samples as negative samples $t_{neg}$. Using these query-product relevance data, we use the pairwise loss as the training objective:
\begin{equation}
\mathcal{L}_{rele} = max \left(0, S_{rele}(q, t_{neg}|c) - S_{rele}(q, t_{pos}|c) + \mu \right) \,, \\
\end{equation}
where $\mu$ is the margin of the pairwise loss. 

% \clearpage

\begin{table}[!htbp]
  \caption{Dataset statistics.}
  \centering
  \label{tab:datset}
  \setlength{\tabcolsep}{1mm}{
      \begin{tabular}{c|c|c}
                \toprule
                % \multirow{2}{*}{\textbf{Models}}     \\
                % &\multicolumn{4}{c}{\textbf{Intent Data}} \\
                \textbf{Statistics} &\textbf{SFT Dataset} &\textbf{Alignment Dataset} \\
                \midrule
                \midrule
                \#Query-Code Pairs  & 184.8M & 67.4M \\
                \hline 
                \#Product-Code Pairs  & 459.7M & 134.8M \\
                \hline 
                \#Uniq. Queries  & 6.2M & 1.5M  \\
                \hline
                \#Uniq. Products  & 8.4M & 15.6M \\
                \hline
                \#Uniq. Codes  & 7.4M & 452.7K \\
                \hline
                Avg. \#chars of query & 8.0 & 7.0 \\
                \hline
                Avg. \#chars of product & 50.3 & 51.9 \\
                \hline
                Avg. \#chars of code & 9.9 & 7.9 \\
                \bottomrule
        \end{tabular}
    }
\end{table}

\section{Experiment}
This section will discuss the offline and online experiments in detail. We first introduce the datasets and the evaluation metrics used in this paper. Then, we analyze the experiment results by several fair comparisons with strong baselines. After that, we deeply investigate the effect of different modules of the GRAM model. Subsequently, we present the online performance of the model on the JD search engine and further analyze the influence of various modules. Finally, we explore the influence of hyper-parameters.

\subsection{Dataset}
\label{sec:Dataset}
To evaluate the effectiveness and generality of the proposed model, we conducted a series of experiments on two large-scale real-world datasets collected from users' click logs on the JD application. The statistics of the datasets are listed in Table~\ref{tab:datset}. Specifically, 
\begin{itemize}
    \item \textbf{SFT Dataset}: Based on online click data, we collected query and product pairs from the past month. Using the results of named entity recognition (NER), we constructed initial query-code and product-code associations. Queries and products that co-occurred were linked to their respective 184.8M query-code pairs and 459.7M product-code pairs. The dataset includes 6.2 million queries, 7.4 million codes, and 8.4 million products.
    \item \textbf{Alignment Dataset}: From the online impression logs, we extracted click data from the past seven days and highly relevant impression data from the first three pages to construct a query-product pair dataset. Using SFT training data, we obtained query-code and product-code information. For each query-product pair, we identified the intersection of their tags as the positive example set and the difference as the negative example set, thereby constructing a partially ordered dataset.
\end{itemize}

\subsection{Baseline Models}
We compare GRAM with several strong baseline models, including widely used sparse retrieval methods, dense retrieval methods, and generative retrieval methods. The detailed introductions are listed as follows:

\begin{itemize}
    \item \textbf{BM25}~\cite{robertson2009probabilistic}: It enhances relevance scoring by refining term weights based on the TF-IDF feature.
    \item \textbf{DocT5Query}~\cite{nogueira2019doc2query}: It utilizes T5 to generate a pseudo query for the document to expand document information and then applies BM25 for document retrieval. 
    \item \textbf{DPR}~\cite{karpukhin2020dense}: It uses BERT as an encoder to encode queries and documents into semantic space and train models with in-batch negatives. 
    \item \textbf{SEAL}~\cite{bevilacqua2022autoregressive}: It regards all n-grams contained in documents as their identifiers. 
    \item \textbf{LC-Rec}~\cite{zheng2024adapting}: It utilizes the RQ-VAE~\cite{zeghidour2021soundstream} to generate semantic IDs for product indexing and proposes a series of semantic alignment tasks to align LLM with semantic IDs.
\end{itemize}

\begin{table}[!htbp]
  \caption{
    The experimental results are compared with sparse retrieval, dense retrieval, and generative retrieval methods. 
  }
  \centering
  \label{tab:experiment}
  \setlength{\tabcolsep}{1mm}{
      \begin{tabular}{c|cccc}
                \toprule
                \textbf{Models} &\textbf{Recall@10} &\textbf{Recall@100} &\textbf{Recall@300} &\textbf{RelR} \\
                \midrule
                \midrule
                \textbf{BM25}  & 3.01\% & 10.52\%  & 15.23\%  & 35.78\%  \\
                \textbf{DocT5Query} & 3.13\% & 10.88\%  & 15.88\%  & 35.56\%  \\
                \textbf{DPR}         & 3.89\% & 11.26\%  & 17.92\%  & 30.96\%  \\
                \midrule 
                \textbf{SEAL}  & 3.25\% & 11.62\%  & 16.56\%  & 27.03\%  \\
                \textbf{LC-Rec}      & 4.35\% & 7.16\%  & 7.33\%  & 23.94\%  \\
                \midrule
                \textbf{GRAM}  & 2.85\% & 12.54\%  & 21.13\%  & 40.18\%  \\
                \bottomrule
        \end{tabular}
    }
\end{table}

\subsection{Evaluation Metrics}
Recall@k is a metric that measures the proportion of expected documents retrieved by the search system. For a given cutoff point k, Recall@k is defined as:
\begin{equation}
    Recall@k = \frac{1}{|Q|} \sum^{|Q|}_{q=1} \frac{ret_{q,k}}{rel_q} \,,
\end{equation}
where $|Q|$ is the number of queries in the set, $ret_{q,k}$ is the number of expected documents retrieved for the $q$-th query within the top k results, and $rel_q$ is the total number of expected documents for the $q$-th query. The expected documents for one query are documents that users have clicked.

The relevance ratio (RelR) is a metric that measures the proportion of relevant documents retrieved by the search system. RelR differs from recall@k in that the relevance ratio does not rely on post-hoc user clicks, allowing for a more comprehensive evaluation of a method's effectiveness from a relevance perspective.

\subsection{Experiment Settings}
For LLM fine-tuning and alignment, we implement the models based on the Pytorch framework and utilize an internal $0.5$B LLM to generate query code and product code. We use the AdamW optimizer with a learning rate of $5e^{-5}$. The max length of the query is set to 16 and the maximum number of codes is set to 10, and each code can contain up to 6 attributes. During the Supervised Fine-Tuning phase, the weight parameter $\lambda$ in the loss function is set to 1.

We use the dropout strategy with a dropout rate of 0.05 to overcome overfitting. The maximum training epoch is set to 3, and the batch size of the training set is set to 128. We select the best parameter configuration based on the performance of the validation set and evaluate the configuration on the test set.

\subsection{Offline Evaluation}

\subsubsection{Offline performance}
The experimental results are shown in Table~\ref{tab:experiment}. Overall, the experimental results indicate that GRAM significantly outperforms all baselines on a large-scale real-world dataset. Specifically, we have the following observations:

(1) Compared with the sparse and dense retrieval methods (i.e., BM25, DocT5Query, DPR), it is obvious that GRAM outperforms them by a significant margin on the dataset. Sparse retrieval methods mainly focus on character-granular matching. It performs well for frequently searched queries. However, they still struggle with intricate queries involving synonyms, and specialized terms, underscoring the need for ongoing advancements to meet users' diverse information needs better. Dense retrieval is limited by the construction of positive and negative samples, and general world knowledge cannot be used. 

(2) Compared with generative retrieval methods, GRAM demonstrates superior effectiveness and practicality. SEAL utilizes all n-grams as codes without achieving alignment between queries and SKUs, which results in reduced retrieval efficiency and lower recall rates. The LC-Rec method, on the other hand, employs semantic IDs that necessitate aligning general language with these IDs from scratch, heavily depending on training data and potentially diminishing the relevance ratio. Additionally, the complexity of generating LC-Rec is linear concerning the number of recalled products. In contrast, the GRAM method effectively integrates the capabilities of general language understanding by aligning queries and products through shared code. It further enhances performance by aligning with relevance metrics specific to the retrieval domain, thereby achieving superior results.

In conclusion, GRAM demonstrates a substantial improvement over all baseline models in terms of Recall@k, and relevance. The results confirm that GRAM can enhance retrieval efficiency and query-product relevance simultaneously.

\begin{table}[!htbp]
  \caption{
    Ablation study of GRAM. (1) Co-training module (CT): This component aligns code generation on both the query and product sides. (2) Co-alignment module (CA): This component aligns offline generation with online relevance.
  }
  \centering
  \label{ablation_study}
  \setlength{\tabcolsep}{1mm}{
      \begin{tabular}{c|cccc}
                \toprule
                \textbf{Models} &\textbf{Recall@10} &\textbf{Recall@100} &\textbf{Recall@300} &\textbf{RelR} \\
                \midrule
                \textbf{GRAM}  & 2.85\% & 12.54\%  & 21.13\%  & 40.18\%  \\
                \midrule
                \textbf{w/o. CA}  & 1.80\% & 11.37\%  & 20.23\%  & 33.51\%  \\
                \textbf{w/o. CT\&CA}  & 1.57\% & 7.64\%  & 12.89\%  & 33.36\%  \\
                \bottomrule
        \end{tabular}
    }
\end{table}

\subsubsection{Ablation study}
To further figure out the relative importance of each module in the proposed model, we perform a series of ablation studies over the different components of GRAM. Two variants of GRAM are listed below: 

\begin{itemize}
    \item \textbf{w/o Co-alignment operation}: When removing the Co-alignment process, the model exhibits a consistent decline in recall rate, along with a significant decrease in relevance result. This indicates that the co-alignment operation can significantly improve the model's capability to capture similarities between queries and products.
    
    \item \textbf{w/o Co-training \& Co-alignment operation}: When we eliminate both Co-training and Co-alignment operation, the performance degrades by more than 8\% in recall@300 compared with the complete GRAM. The results indicate that improving the interaction between query-code and product-code can significantly enhance retrieval accuracy.
\end{itemize}

\subsection{Online Evaluation}

\subsubsection{Online Deployment}
In online platforms for search advertising, the process is primarily divided into two components: retrieval and ranking. The online system already includes retrieval branches such as sparse retrieval methods like BM25 and DocT5query, as well as dense retrieval methods like DPR. 

In this study, we introduce a generative retrieval method in the retrieval phase. To mitigate latency issues, we utilize a real-time cache to reduce the number of inferences, and we set the beam search size to 10, ultimately returning 300 advertisements. Compared to the semantic ID generation approach, this method generates results at a scale of an order of magnitude smaller, making it more suitable for large-scale industrial applications.

\begin{figure}[t]
\centering
\includegraphics[width=\columnwidth]{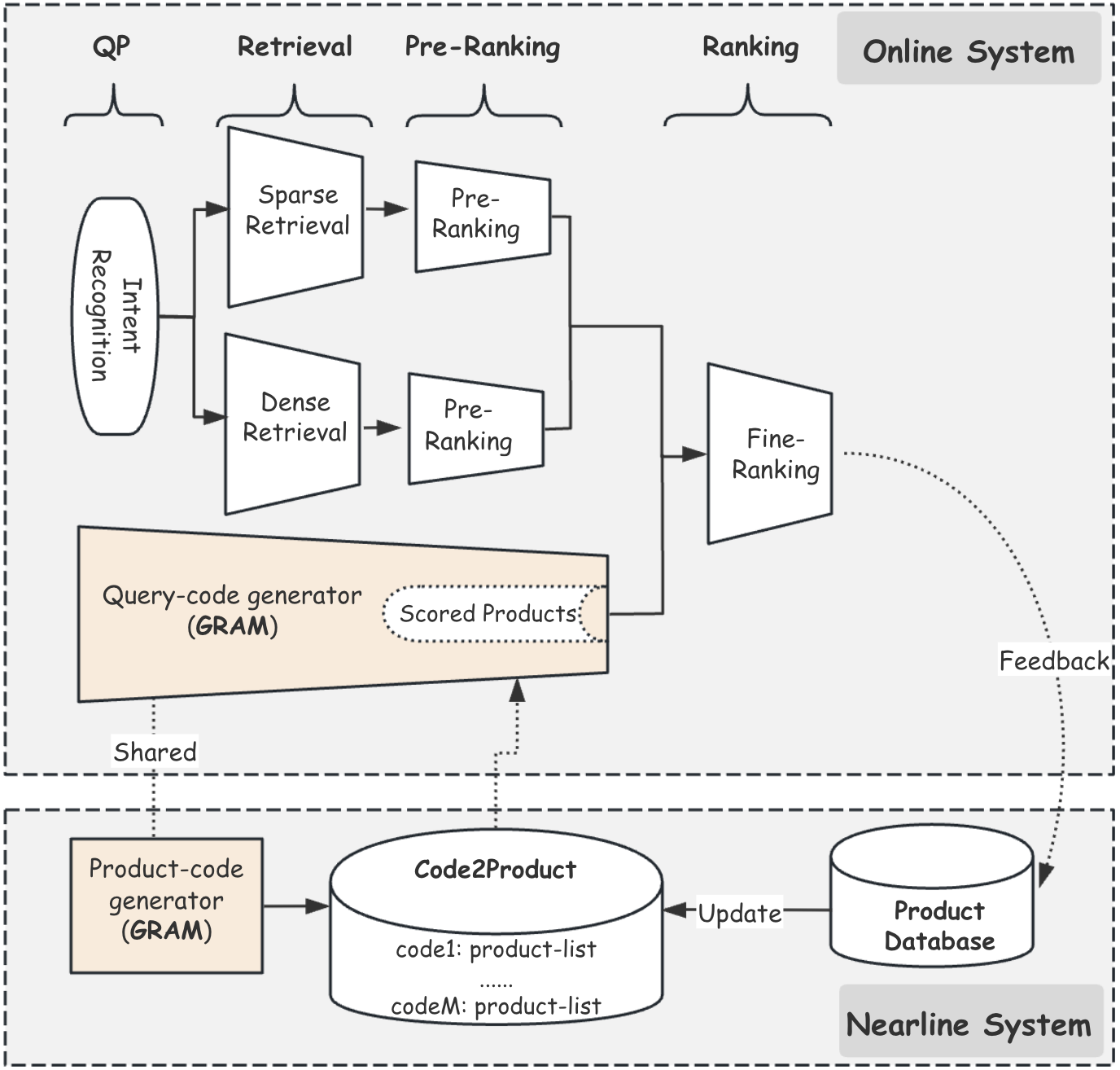}
\caption{The deployment of GRAM and the role of retrieval plays in the E-commerce system.}
\label{fig:system}
\end{figure}

Figure~\ref{fig:system} illustrates the role of GRAM within the system. GRAM effectively merges the stages of intent recognition, retrieval, and pre-ranking, thereby reducing the information loss typically associated with the layered funnel approach. The same GRAM model operates in both online and nearline systems. Initially, GRAM generates relevant codes based on queries. These codes are then used to ultimately produce items with associated scores. In the nearline component, it generates codes for new products in response to changes in inventory and updates these codes in the product index.

\subsubsection{Online Performance}
Before being launched in production, we routinely deployed the GRAM online on the JD search engine and made it randomly serve 5\% traffic as the test group. For a fair comparison, we also built a base group that uses the previous model to serve 5\% traffic. During the A/B testing period, we monitor the performance of GRAM and compare it with the online model. This period lasts for at least one week.

For online evaluation, we use some business metrics: Ad. imp.. (total count of ad impressions), CPC (cost per click), CTR (click-through rate), and Ad. revenue(total ad revenue). The specific computation of CPC and CTR is defined as follows: 
\begin{equation}
\begin{split}
    & CPC = \frac{Ad. revenue}{\#Clicks} \,, \\
    & CTR = \frac{\#Clicks}{\#Impressions}   \,,  \\
\end{split}
\end{equation}
where \#Clicks represents the total number of user clicks on an advertisement, while \#Impressions denote the total number of times the advertisement is displayed to users. 

\begin{table}[!htbp]
  \centering
  \caption{Online improvements of the GRAM. Improvements are statistically significant with $p < 0.05$ on a paired t-test. All performances of GRAM and its variants GRAM without Co-training (CT) and without Co-alignment (CA) are compared with the online model.}
  \label{online_performance_uv_ucvr}
  \setlength{\tabcolsep}{1mm}{
      \begin{tabular}{c|cccc}
            \toprule
            \textbf{Models} &\textbf{Ad. imp.}  &\textbf{CTR} &\textbf{CPC}  & \textbf{Ad. revenue}
            \\
            \midrule
            \textbf{Online}          & - & -  & - & -   \\
            \textbf{GRAM w/o. CT\&CA}   & +0.37\%   & +0.21\%  & +0.79\%    & +1.37\%   \\
            \textbf{GRAM w/o. CA}   & +0.56\%   & +0.92\%  & +0.68\%    & +2.16\%   \\
            \textbf{GRAM}            & +0.74\%   & +1.27\%  & +0.45\%    & +2.46\%   \\
            \bottomrule
        \end{tabular}
    }
\end{table}

Compared to the existing retrieval branches in the online system, such as sparse retrieval methods like BM25 and DocT5query, and dense retrieval methods like DPR, GRAM has delivered significant improvements in advertising revenue. It has achieved substantial gains in click-through rate (CTR), cost per click (CPC), and overall ad revenue, validating the effectiveness of GRAM. The experimental results comparing GRAM and its variant models align with the ablation study, demonstrating the effectiveness of dual alignment: the alignment of codes between queries and products, and the alignment of code generation with online relevance.

\section{Conclusion and Future Work}
In this paper, we have introduced the Generative Retrieval and Alignment Model (GRAM), a novel approach to e-commerce retrieval that leverages the capabilities of large language models (LLMs) to generate shared codes for both queries and products. By effectively addressing the challenges of inconsistent code generation due to differing word distributions, GRAM enhances retrieval efficiency and accuracy. Our approach integrates retrieval and pre-ranking processes, significantly outperforming traditional sparse and dense retrieval methods, as well as existing generative retrieval models. The deployment of GRAM on the JD e-commerce platform, where it handles millions of product retrievals daily, underscores its commercial viability and scalability.

For future work, we aim to further unleash the potential of large language models by incorporating multi-modal and personalized information into the GRAM framework. Building on the integration of retrieval and pre-ranking, a promising goal is to develop an end-to-end search system that encompasses the entire decision-making chain.

\clearpage

%%
%% The next two lines define the bibliography style to be used, and
%% the bibliography file.
\begin{spacing}{1.35}
\bibliographystyle{ACM-Reference-Format}
\balance
\bibliography{sample-sigconf}
\end{spacing}

%%
%% If your work has an appendix, this is the place to put it.
% \appendix

% \section{Research Methods}

% \subsection{Part One}

\end{document}